\documentclass[aps,longbibliography,showpacs,floatfix,twocolumn]{revtex4-2} 
\usepackage{graphicx}

\usepackage{times,bm,url,amsmath,upgreek}
\usepackage{amssymb}
\usepackage[colorlinks=true,breaklinks=true,allcolors=blue]{hyperref}
\usepackage[utf8]{inputenc}

\usepackage{comment}
\usepackage{subfigure}
\usepackage[scr=boondox]{mathalfa}
\usepackage{xparse}
\usepackage{bbm}
\usepackage{cases}
\usepackage{bbold} 

\begin{document}

\title{Pole-skipping in the de Sitter horizon structure}

\author{Haiming Yuan}
\email[Corresponding author. ]{yuan\_haiming@haut.edu.cn}
\affiliation{School of Physics and Advanced Energy, Henan University of Technology, 100 Lianhua Street, Zhengzhou 450001, China}
\author{Xian-Hui Ge}
\affiliation{Department of Physics, College of Sciences, Shanghai University, 99 Shangda Road, Shanghai 200444, China}
\author{Keun-Young Kim}
\affiliation{Department of Physics and Photon Science, Gwangju Institute of Science and Technology, 123 Cheomdan-gwagiro, Gwangju 61005, Korea}
\affiliation{Research Center for Photon Science Technology, Gwangju Institute of Science and Technology, 123 Cheomdan-gwagiro, Gwangju 61005, Korea}

\begin{abstract}
We study the pole-skipping structure of incoming waves near the cosmic horizon $r=r_c$ in de Sitter (dS) spacetime. We find that the scalar field with spin-0, the Dirac field with spin-1/2, the Maxwell field with spin-1, the Rarita-Schwinger field with spin-3/2, and the gravitational field with spin-2 exhibit the same frequencies $\omega_\star$ of pole-skipping points as their corresponding spin fields satisfying the incoming wave conditions in anti-de Sitter (AdS) spacetime. However, the momenta $k_\star$ undergo a shift in complex space, which originates from the spacetime curvature. The shift in momenta at the pole-skipping points could be measured through the operator dimensions $\Delta$ in two curved spacetimes, where momenta are mutually complex conjugate in dS and AdS, two opposite curvature spacetimes.
\end{abstract}

\maketitle
\section{Introduction}
\label{sec:intro}
``Pole-skipping" is a phenomenon with very interesting properties in the anti-de Sitter/conformal field theory (AdS/CFT) theory. Generally, the retarded Green's function takes the form
\begin{equation}
\label{eq:1}
G^R(\omega,k)=\frac{b(\omega,k)}{a(\omega,k)}
\end{equation}
in the complex momentum space $(\omega,k)$. At a special point $(\omega_\star, k_\star)$ both $a$ and $b$ satisfy $a(\omega_\star, k_\star)=b(\omega_\star, k_\star)=0$, and the retarded Green's function cannot be uniquely defined~\cite{Grozdanov1,Blake1,Blake2,Grozdanov2,Das}. Its value will be determined by how it approaches this special point, that is, it depends on the slope $\delta k/\delta \omega$.
\begin{equation}
\label{eq:2}
G^R=\frac{(\partial_\omega b)_\star +\frac{\delta k}{\delta \omega}(\partial_k b)_\star+\dots}{(\partial_\omega a)_\star +\frac{\delta k}{\delta \omega}(\partial_k a)_\star+\dots}.
\end{equation}
So if we find the intersections of zeros and poles in the retarded Green's functions, we can obtain those special points, which we refer to as pole-skipping points. For the theory of the AdS/CFT correspondence, we can use another method to obtain the pole-skipping points from the bulk field equation \cite{Makoto1,Makoto2,BlakeDavison,Abbasi1,Abbasi2,Abbasi3,Choi,Karunava,Ahn1,Mahdi}. The absence of a unique incoming mode near the horizon corresponds to the non-uniqueness of the Green's function on the boundary. For the static black holes in AdS spacetime, the leading-order pole-skipping frequency is known as $\omega=i2\pi T(s-1)$~\cite{Makoto3,N1,Yuan1,N2,Yuan2,Diandian,Jeong,Yuan3,Ning,Ahn2}, where $i$ is the imaginary unit, and $s$ denotes the spin of the operator.

Recently, people have begun to study the pole-skipping structure in de Sitter (dS) spacetime~\cite{Grozdanov3,Yuan4,Ahn3}, attempting to find similarities and differences between it and this special structure in AdS spacetime. According to these results, it can be obtained that the frequency of the pole-skipping point is related to the selection of incoming and outgoing conditions near the horizon. Selecting different Eddington-Finkelstein (EF) coordinate conditions, $u=t-r_*$ or $v=t+r_*$, at the dS and AdS horizons leads to a frequency transition from $\omega$ to $-\omega$~\cite{Grozdanov3,Yuan4,Ahn3}. The frequencies of the leading order pole-skipping locations of bosonic fields, such as the scalar field (spin-0), the vector field (spin-1), and the gravitational field (spin-2), that satisfy the incoming wave condition near the dS horizon are the same as the frequencies of the corresponding spin fields that also satisfy the incoming wave condition near the AdS horizon, however the momentum $k$ between them is different~\cite{Ahn3}. We hope to further consider more general situations and analyze the patterns and underlying reasons for frequency and momentum distributions across different spacetime configurations.

In this paper, we further explore the higher-order pole-skipping structure of the bosonic field, and newly investigate the pole-skipping structure of the fermionic field in high-dimensional dS spacetime. The locations of the pole-skipping points in the higher-dimensional dS spacetime are located in the space $(\omega, k)$, rather than in space $(\omega, m)$ as in the two-dimensional dS spacetime case. In two-dimensional dS spacetime, the mass $m$ of the pole-skipping point is the complex conjugate of the mass of the corresponding special point in two-dimensional AdS~\cite{Yuan3,Yuan4}. Interestingly, a similar phenomenon occurs in higher-dimensional spacetimes. If we consider the incoming wave condition in dS spacetime, the frequency of the pole-skipping point is the same as the corresponding result in AdS spacetime, while the opposite curvature leads to the opposite momentum $k$ value on the imaginary axis in terms of the operator dimension $\Delta$.

We calculate the higher-order pole-skipping points for the bosonic fields in de Sitter spacetime in Sec.~\ref{sec:B}. Then we study the pole-skipping points for the fermionic fields in de Sitter spacetime in Sec.~\ref{sec:F}. Finally, we summarize and discuss in Sec.~\ref{sec:Conclusion}.

\section{Bosonic field}
\label{sec:B}
The leading-order pole-skipping results for the scalar field, the Maxwell field, and the gravitational field in dS spacetime are all presented in Ref.~\cite{Ahn3}. We could find that the frequencies of their pole-skipping points are the same, but the momenta will appear differently in complex space if using the incoming wave conditions at both the AdS and dS horizons~\cite{Ahn3}. In this section, we will continue to solve those fields for higher-order pole-skipping points and compare the results with those at the AdS horizon, and we will obtain that these differences in momenta come from the different curvatures of geometric structures.
\subsection{Scalar field}
\label{sec:1}
We consider the scalar field with spin-0 in de Sitter spacetime in this subsection. The metric of de Sitter spacetime in static coordinates is given as~\cite{Gibbons,Strominger1}
\begin{equation}
	\label{eq:1.1}
	ds^2=-f(r)dt^2+\frac{1}{f(r)}dr^2+h(r)d\Omega^2_{d-2},
\end{equation}
where $f(r)=1-r^2/L^2$ and $L$ is the radius of the dS. The cosmic horizon at $r_c=L$, and the global temperature is given by $T_{dS}=\frac{1}{2\pi L}$~\cite{Gibbons,Strominger1,Strominger2,Witten3}. Since a single static patch in the dS spacetime is located in the region where $r<r_c$, the incoming wave condition at $r_c$ may satisfy $dr_*/dt>0$~\cite{Ahn3}. Therefore we use the incoming Eddington-Finkelstein (EF) coordinate $u=t-r_*$, where $r_*$ is the tortoise coordinate $dr_*=dr/f(r)$, and ~\eqref{eq:1.1} becomes
\begin{equation}
	\label{eq:1.2}
	ds^2=-f(r)du^2-2dudr+h(r)dx^2_{\alpha}.
\end{equation}
We use the symbol $x_{\alpha}$ to label the $(d-2)-$dimensional space $\alpha=1,2,\dots$. We consider the minimally coupled massive scalar field $\Phi(u,x,r)=e^{-i \omega u+i\vec{k}\cdot\vec{x}}\Phi(r)$ which the dynamics are given by the Klein-Gordon equation
\begin{equation}
\label{eq:1.3}
\frac{1}{\sqrt{-g}}\partial_\mu(\sqrt{-g}g^{\mu\nu}\partial_\nu\Phi)-m^2\Phi=0\,.
\end{equation} 
It could be expanded as
\begin{equation}
\label{eq:1.4}
\begin{aligned}
&h(r)f(r)\Phi''(r)+\big((d-2)h'(r)f(r)+h(r)f'(r)+2i\omega h(r)\big)\\
&\times\Phi'(r)+\big(\frac{d-2}{2}i\omega h'(r)-k^2-m^2h(r)\big)\Phi(r)=0.
\end{aligned}
\end{equation}
Since the blacken factor goes $f(r) = 4\pi T(r_c-r) + \mathcal{O}\left(r_c-r\right)^2$ near the horizon, one can check that the Klein-Gordon equation~\eqref{eq:1.4} has a regular singularity at $r=r_c$. The singularity can be seen if we approximate~\eqref{eq:1.4} to
\begin{equation}
\begin{aligned}
\label{eq:1.5}
&\Phi''(r)+\frac{\frac{i\omega}{2\pi T}-1}{r_c-r}\Phi'(r)\\
&+\frac{1}{4\pi T h(r_c)}\frac{\frac{d-2}{2}i\omega h'(r_c)-k^2-m^2h(r_c)}{r_c-r}\Phi(r)=0,
\end{aligned}
\end{equation}
near the horizon. According to conventional differential equation techniques, one can solve~\eqref{eq:1.5} by imposing series solution ansatz as
\begin{equation}
\label{eq:1.6}
\Phi(r)=(r-r_c)^\lambda\sum^\infty_{p=0}\Phi_p(r-r_c)^p.
\end{equation}
At the lowest order, we can obtain the indicial equation $\lambda(\lambda-\frac{i\omega}{2\pi T})=0$. The two roots are
\begin{equation}
\label{eq:1.7}
\lambda_1=0,\quad \lambda_2=\frac{i\omega}{2\pi T}\,.
\end{equation}
Generally speaking, a solution with exponent $\lambda_1$ near the horizon is regular. Therefore, if $\lambda_2$ is not an integer, we can obtain a unique ``incoming" solution with exponent $\lambda_1$. However, if $\lambda_2$ is an integer, there may be an additional regular solution, which is a signal of a pole-skipping near the horizon. For example, let us consider $i\omega=2\pi T$. According to the standard technique of the differential equation, the assumption for the following solution is more appropriate than~\eqref{eq:1.6}
\begin{equation}
\begin{aligned}
\label{eq:1.8}
\Phi(r)=&\sum^\infty_{p=0}\Phi_{1,p}(r-r_c)^p\\
&+(r-r_c)\log(r-r_c)\sum^\infty_{q=0}\Phi_{2,q}(r-r_c)^q\,,
\end{aligned}
\end{equation}
where we include $\log$ term. After substituting it into the Klein-Gordon equation~\eqref{eq:1.4}, up to $\mathcal{O}(r-r_c)^0$ the equation of motion becomes
\begin{equation}
\begin{aligned}
\label{eq:1.9}
&\big(1-\frac{i\omega}{2\pi T}\big)\big(\Phi_{1,1}+ \Phi_{2,0}\log(r-r_c)\big)+\big(\big(2-\frac{i\omega}{2\pi T}\big)\Phi_{2,0}\\
&+\frac{\frac{d-2}{2}i\omega h'(r_c)-k^2-m^2h(r_c)}{4\pi T h(r_c)}\Phi_{1,0}\big)=0\,.
\end{aligned}
\end{equation}
The first term vanishes at $i\omega=2\pi T$. If we focus on $(\frac{d-2}{2}i\omega h'(r_c)-k^2-m^2h(r_c))=0$, we obtain $\Phi_{2,0}=0$, and the series solution takes two independent regular solutions as follows
\begin{equation}
\label{eq:1.10}
\Phi(r)=\Phi_{1,0} + (r-r_c)^1\sum^\infty_{p=0}\tilde{\Phi}_{2,p}(r-r_c)^p\,,
\end{equation}
where two independent coefficients are $\Phi_{1,0}$ and $\tilde{\Phi}_{2,0}$. Thus, the leading-order pole-skipping point is
\begin{equation}
\label{eq:1.11}
\omega_{\star}=-i2\pi T,\quad k^2_{\star}=\pi T(d-2)h'(r_c)-m^2h(r_c)\,,
\end{equation}
which is the same as the numerical value in Ref.~\cite{Ahn3}.

Now we consider the higher-order locations. To compare with the results obtained for BTZ black holes in AdS, we choose $d=3$ and $h(r)=r^2$. In general, at the pole-skipping points, the series solution takes the following form
\begin{equation}
\label{eq:1.12}
\Phi(r)=\sum^{n-1}_{p=0}\Phi_{1,p}(r-r_c)^p + (r-r_c)^n\sum^\infty_{q=0}\tilde{\Phi}_{2,q}(r-r_c)^q\,,
\end{equation}
where $\Phi_{1,p}$ and $\tilde{\Phi}_{2,q}$ are independent coefficients. There is a systematic procedure to calculate pole-skipping points~\cite{BlakeDavison}. Firstly, we expand $\Phi(r)$ with a Taylor series
\begin{equation}
\label{eq:1.13}
\Phi(r)=\sum^\infty_{p=0}\Phi_p(r-r_c)^p=\Phi_0+\Phi_1(r-r_c)+\Phi_2(r-r_c)^2+\dots.
\end{equation}
We substitute \eqref{eq:1.13} into \eqref{eq:1.4} and expand the equation of motion in powers of $(r-r_c)$. Then, a series of the perturbed equation in the order of $(r-r_c)$ can be denoted as
\begin{equation}
\label{eq:1.14}
S=\sum^\infty_{p=0}S_p(r-r_c)^p=S_0+S_1(r-r_c)+S_2(r-r_c)^2+\cdots\,.
\end{equation}
We write down the first few equations $S_p=0$ in the expansion of~\eqref{eq:1.14}:
\begin{equation}
\begin{aligned}
0=&M_{11}(\omega,k)\Phi_0+(2\pi T-i\omega)\Phi_1,\\
0=&M_{21}(\omega,k)\Phi_0+M_{22}(\omega,k)\Phi_1+(4\pi T-i\omega)\Phi_2,\\
0=&M_{31}(\omega,k)\Phi_0+M_{32}(\omega,k)\Phi_1+M_{33}(\omega,k)\Phi_2\\
&+(6\pi T-i\omega)\Phi_3\,.
\end{aligned}
\end{equation}
To find an incoming solution, we should solve a set of linear equations of the form
\begin{equation}
   		\begin{aligned}
   \label{eq:1.15}
   &\mathcal{M}(\omega,k)\cdot \Phi\equiv\\
   &\left(\begin{array}{ccccc}
   	M_{11} & (2\pi T-i\omega) & 0    & 0  &\dots\\
   	M_{21} & M_{22}& (4\pi T-i\omega)& 0   &\dots\\
   	M_{31} & M_{32}&  M_{33} &(6\pi T-i\omega) &\dots\\
   	\dots   &  \dots&  \dots  &\dots   &\dots\\
   \end{array}\right)\\
   &\times\left(\begin{array}{ccccc}
   	\Phi_0\\
   	\Phi_1\\
   	\Phi_2 \\
   	\dots \\
   \end{array}\right)=0\,.
   	\end{aligned}
   \end{equation}
The $n$-th pole-skipping points $(\omega_{n}, k_{n})$ can be calculated by solving
\begin{equation}
\label{eq:1.16}
\omega_{\star n}=i2\pi Tn,\qquad {\rm det}\mathcal{M}^{(n)}(\omega_\star,k_\star)=0\,,
\end{equation}
where the matrix $\mathcal{M}^{(n)}$ is the $(n\times n)$ square matrix whose elements are taken from $M_{11}$ to $M_{nn}$ in~\eqref{eq:1.15}. (The first few elements of this matrix have been shown in Appendix~\ref{sec:Details1}.) The following are the resulting  all-order pole-skipping points:
\begin{equation}
\label{eq:1.17}
\begin{aligned}
\omega_{\star}=-2i\pi T, \quad k_{\star}=&\pm\sqrt{1-\frac{m^2}{4\pi^2T^2}}\,;\\
\omega_{\star}=-4i\pi T, \quad k_{\star}=&\pm 1\pm\sqrt{1-\frac{m^2}{4\pi^2T^2}}\,;\\
\omega_{\star}=-6i\pi T, \quad k_{\star}=&\pm\sqrt{1-\frac{m^2}{4\pi^2T^2}},\;\pm 2\pm\sqrt{1-\frac{m^2}{4\pi^2T^2}}\,;\\
\omega_{\star}=-8i\pi T, \quad k_{\star}=&\pm 1\pm\sqrt{1-\frac{m^2}{4\pi^2T^2}},\\
 &\pm 3\pm\sqrt{1-\frac{m^2}{4\pi^2T^2}}\,;\\
\omega_{\star}=-10i\pi T, \quad k_{\star}=&\pm\sqrt{1-\frac{m^2}{4\pi^2T^2}},\;\pm 2\pm\sqrt{1-\frac{m^2}{4\pi^2T^2}},\\
&\pm 4\pm\sqrt{1-\frac{m^2}{4\pi^2T^2}}\,;\\
&\qquad \qquad \vdots
\end{aligned}
\end{equation}
The operator dimension of the scalar field in de Sitter spacetime is $\Delta=\frac{1}{2}\big((d-1)\pm\sqrt{(d-1)^2-4m^2L^2}\big)$~\cite{Strominger1,Strominger2,Witten3}. The pole-skipping points~\eqref{eq:1.17} can be rewritten in terms of operator dimensions:
\begin{equation}
\label{eq:1.18}
\begin{aligned}
\omega_{\star} =-2i\pi T, \quad k_{\star}=&\,\Delta-1\,;\\
\omega_{\star} =-4i\pi T, \quad k_{\star}=&\,\Delta-2,\;\Delta\,;\\
\omega_{\star} =-6i\pi T, \quad k_{\star}=&\,\Delta-3,\;\Delta-1,\;\Delta+1\,;\\
\omega_{\star} =-8i\pi T, \quad k_{\star}=&\,\Delta-4,\;\Delta-2,\;\Delta,\;\Delta+2\,;\\
\omega_{\star} =-10i\pi T, \quad k_{\star}=&\,\Delta-5,\;\Delta-3,\;\Delta-1,\;\Delta+1,\;\Delta+3\,;\\
&\qquad \qquad \vdots
\end{aligned}
\end{equation}
The pole-skipping of the scalar field for BTZ black holes in AdS is given as~\cite{BlakeDavison}:
\begin{equation}
\label{eq:1.19}
\begin{aligned}
\omega_{\star} =-2i\pi T, \quad k_{\star}=&\,i(\Delta-1)\,;\\
\omega_{\star} =-4i\pi T, \quad k_{\star}=&\,i(\Delta-2),\;i\Delta\,;\\
\omega_{\star} =-6i\pi T, \quad k_{\star}=&\,i(\Delta-3),\;i(\Delta-1),\;i(\Delta+1)\,;\\
\omega_{\star} =-8i\pi T, \quad k_{\star}=&\,i(\Delta-4),\;i(\Delta-2),\;i\Delta,\;i(\Delta+2)\,;\\
\omega_{\star} =-10i\pi T, \quad k_{\star}=&\,i(\Delta-5),\;i(\Delta-3),\;i(\Delta-1),\\
&i(\Delta+1),\;i(\Delta+3)\,;\\
&\qquad \qquad \vdots
\end{aligned}
\end{equation}
We find that in comparison between Eq.~\eqref{eq:1.18} and Eq.~\eqref{eq:1.19}, the frequencies are the same, but the momenta are complex conjugates of each other in terms of the operator dimension. The reason for the change in momenta is that the spacetime curvature of dS and AdS is opposite, which mathematically leads to the mass squared $m^2$ and momentum squared $k^2$ being mutually negative in the equation of motion between the two spacetimes. The inverse of mass squared $m^2$ causes $\Delta(\Delta+1-d)=m^2L^2$ in AdS to become $\Delta(\Delta+1-d)=-m^2L^2$ in dS, which means the masses of the stable scalar make the conformal weights become complex, hence the boundary CFT is non-unitary. Meanwhile, the inversion of momentum squared $k^2$ in the equation of motion results in the momenta of the pole-skipping points being opposite on the imaginary axis in terms of the operator dimension.

\subsection{Maxwell field}
\label{sec:2}
We consider the Maxwell field with spin-1 in dS$_4$ spacetime in this subsection. The metric is also described in~\eqref{eq:1.2}. We assume that perturbations take the plane-wave form $A_M\propto e^{-i \omega u+ikx}A_M(r)$. The waves $A_M$ have four components $(M=u,r,x,y)$, and the perturbations are decomposed as
\begin{equation}
\label{eq:2.1}
\left\{
\begin{aligned}
&{\rm scalar:}\; A_u,\; A_r,\; A_{x},\\
&{\rm vector:}\; A_y.
\end{aligned}
\right.
\end{equation}
We could combine these scalars into the gauge-invariant variables:
\begin{equation}
\label{eq:2.2}
\left\{
\begin{aligned}
	&\mathfrak{A}_u=A_u+\frac{\omega}{k}A_x,\\
	&\mathfrak{A}_r=A_r-\frac{1}{ik}A'_x.
\end{aligned}
\right.
\end{equation}
Therefore, the components of the Maxwell equation 
\begin{equation}
\label{eq:2.3}
\frac{1}{\sqrt{-g}}\partial_\nu\big(\sqrt{-g}g^{\mu\rho}g^{\nu\sigma}(\partial_\rho A_\sigma-\partial_\sigma A_\rho)\big)=0
\end{equation} 
could be decomposed into the scalar part
\begin{equation}
\label{eq:2.4}
\left\{
\begin{aligned}
	&\big(k^2-i\omega h'(r)\big)\,\mathfrak{A}_r-h'(r)\,\mathfrak{A}'_u-h(r)\big(i\omega \,\mathfrak{A}'_r+\,\mathfrak{A}''_u\big)=0, \\
	&\big(\omega^2h(r)-k^2f(r)\big)\,\mathfrak{A}_r+k^2\,\mathfrak{A}_u-i\omega h(r)\,\mathfrak{A}'_u=0,
\end{aligned}
\right.
\end{equation}
and the vector part 
\begin{equation}
\label{eq:2.5}
-k^2\,A_y+h(r)\big(f'(r)+2i\omega\big)\,A'_y+h(r)f(r)\,A'_y=0.
\end{equation} 
We can obtain the values of the pole-skipping points from the system of equations~\eqref{eq:4.4} in the scalar part:
\begin{equation}
\label{eq:2.6}
\omega_{\star}=0, \quad k_{\star}=0.
\end{equation}
Meanwhile, we can use the method introduced in the previous two sections to obtain a series of higher-order pole-skipping point values from the equation~\eqref{eq:2.5} in the vector part. (The first few elements of the matrix $\mathcal{M}^{(n)}$ in Maxwell equation, similar to those in matrix equation~\eqref{eq:1.15}, are presented in Appendix~\ref{sec:Details2}.) We use the $h(r)=r^2$ for convenience, and obtain:
\begin{equation}
\label{eq:2.7}
\begin{aligned}
	\omega_{\star}=-2i\pi T, \quad k_{\star}&=0\,;\\
	\omega_{\star}=-4i\pi T, \quad k_{\star}&=0,\;\pm\sqrt{2}\,;\\
	\omega_{\star}=-6i\pi T, \quad k_{\star}&=0,\;\pm\sqrt{2},\;\pm\sqrt{6}\,;\\
	\omega_{\star}=-8i\pi T, \quad k_{\star}&=0,\;\pm\sqrt{2},\;\pm\sqrt{6},\;\pm\sqrt{12}\,;\\
	\omega_{\star}=-10\pi T, \quad k_{\star}&=0,\;\pm\sqrt{2},\;\pm\sqrt{6},\;\pm\sqrt{12},\;\pm\sqrt{20}\,;\\
	&\qquad \qquad \vdots
\end{aligned}
\end{equation}
The all-order pole-skipping points of the Maxwell field in AdS are obtained as:
\begin{equation}
\label{eq:2.8}
\begin{aligned}
&\omega_{\star}=0, \quad\quad\quad\; k_{\star}=0\,;\\
&\omega_{\star}=-2i\pi T, \quad k_{\star}=0\,;\\
&\omega_{\star}=-4i\pi T, \quad k_{\star}=0,\;\pm i\sqrt{2}\,;\\
&\omega_{\star}=-6i\pi T, \quad k_{\star}=0,\;\pm i\sqrt{2},\;\pm i\sqrt{6}\,;\\
&\omega_{\star}=-8i\pi T, \quad k_{\star}=0,\;\pm i\sqrt{2},\;\pm i\sqrt{6},\;\pm i\sqrt{12}\,;\\
&\omega_{\star}=-10\pi T, \quad k_{\star}=0,\;\pm i\sqrt{2},\;\pm i\sqrt{6},\;\pm i\sqrt{12},\;\pm i\sqrt{20}\,;\\
&\qquad \qquad \vdots
\end{aligned}
\end{equation}
Comparing equation~\eqref{eq:2.7} with equation~\eqref{eq:2.8}, we find that the frequencies are the same, but the momenta are opposite on the imaginary axis. The reason is that, similar to the case of the scalar field, the momentum squared $k^2$ in the equation of motion in dS spacetime becomes negative. Unlike in the massive scalar field, due to the masslessness of the Maxwell field, its operator dimension in both AdS spacetime and dS spacetime is the same, $(\Delta-1)(\Delta+2-d)=0$. That is also to say, in terms of the operator dimension, the momenta of~\eqref{eq:2.7} and~\eqref{eq:2.8} are complex conjugates of each other.

\subsection{Gravitational field}
\label{sec:3}
In this subsection, analogous to the Ba$\tilde{\rm n}$ados-Teitelboim-Zanelli (BTZ) black hole~\cite{Banados1,Banados2}, we consider the standard 3-dimensional Einstein-Maxwell theory with a positive cosmological constant. For simplicity, we neglect its coupling with the Maxwell field:
\begin{equation}
	\label{eq:3.1}
	I=\int d^3x\sqrt{-g}\big[R-2\Lambda\big],
\end{equation}
where the cosmological constant $\Lambda$ is related to the radius $L$ by $\Lambda=L^{-2}$. The metric is
\begin{equation}
	\label{eq:3.2}
	ds^2=-f(r)dt^2+\frac{1}{f(r)}dr^2+r^2dx^2,
\end{equation}
where $f(r)=M-r^2/L^2$, and $M$ is a constant of integration, which will be identified below as the mass. This metric has an apparent singularity at $r_c=\sqrt{M}L$, and the global temperature is given by $T_{dS}=\frac{\sqrt{M}}{2\pi L}$. We use the incoming Eddington-Finkelstein (EF) coordinate $u=t-r_*$, where $r_*$ is the tortoise coordinate $dr_*=dr/f(r)$. In the incoming EF coordinate,~\eqref{eq:3.2} becomes
\begin{equation}
	\label{eq:3.3}
	ds^2=-f(r)du^2-2dudr+r^2dx^2.
\end{equation}
The vacuum Einstein equation is given as
\begin{equation}
\label{eq:3.4}
R_{ab}-\frac{1}{2}g_{ab}R+\Lambda g_{ab}=0.
\end{equation} 
We consider the sound modes of the metric perturbation
\begin{equation}
\label{eq:3.5}
\begin{aligned}
&h_{uu}=e^{-i\omega u+ikx}h_{uu}(r),\quad h_{ux}=e^{-i\omega u+ikx}h_{ux}(r),\\
&h_{xx}=e^{-i\omega u+ikx}h_{xx}(r).
\end{aligned}
\end{equation}
We obtain the $uu$ component near the horizon $r=r_c$
\begin{equation}
\label{eq:3.6}
L\big(k^2+i\omega L\sqrt{M}\big)h_{uu}+\big(\omega L-i\sqrt{M}\big)\big(2kh_{ux}+\omega h_{xx}\big)=0.
\end{equation}
From equation~\eqref{eq:3.6}, we can obtain the position of its pole-skipping point:
\begin{equation}
\label{eq:3.7}
\omega_{\star}=2i\pi T, \quad k_{\star}=\sqrt{M}.
\end{equation}
For the BTZ black hole in AdS spacetime with metric
\begin{equation}
	\label{eq:3.8}
	ds^2=-(r^2/L^2-M)dt^2+\frac{1}{(r^2/L^2-M)}dr^2+r^2dx^2,
\end{equation}
the pole-skipping point where the gravitational field equation equals~\eqref{eq:3.4} is:
\begin{equation}
\label{eq:3.9}
\omega_{\star}=2i\pi T, \quad k_{\star}=i\sqrt{M}.
\end{equation}
The gravitational field has no mass, hence the operator dimensions in dS and AdS spacetimes are identical $(\Delta-1)(\Delta+1-d)=0$. Regardless of whether operator dimensions are used as a measure, the momenta at the pole-skipping points in dS and AdS spacetimes are complex conjugates of each other. The frequencies of pole-skipping points on both sides are the same, and this conclusion is consistent with that in Ref.~\cite{Ahn3}.

\section{Fermionic field}
\label{sec:F}
\subsection{Dirac field}
\label{sec:4}
In this section, we consider the Dirac field with spin-1/2 in dS$_3$ spacetime, and as with the scalar field, we will find that the obtained Pole-skipping points also change with the geometric structure of the horizon. The Dirac equation is given as
\begin{equation}
\label{eq:4.3}
(\Gamma^MD_M-m)\psi_{\pm}=0.
\end{equation}
The capital letter $M$ denotes the indices of bulk spacetime coordinates. The covariant derivative of bulk spacetime acting on fermions is defined by $D_M=\partial_M+\frac{1}{4}(\omega_{ab})_M\Gamma^{ab}$, where $\Gamma_{ab}\equiv\frac{1}{2}[\Gamma_a,\Gamma_b]$. $\Gamma_a$ are Gamma matrices which satisfy Grassman algebra $\{\Gamma^a,\Gamma^b\}=2\eta^{ab}$~\cite{Wilczek,Pethybridge}. The small letters $a,b$ denote tangent space indices. The spinors are two dimensional $\psi_{\pm}(u,x,r)=e^{-i\omega u+ikx}\left(
\begin{array}{c}
\psi_{+}(r)\\
\psi_{-}(r)\\
\end{array}
\right)$. The number of components of a spinor is $N=2^{[\frac{d}{2}]}$, where $[q]$ denotes the highest integer that is less than or equal to $q$. The Dirac equation~\eqref{eq:4.3} will become a system of coupled first-order differential equations for the $N$ components of the spinor. In the static coordinate of~\eqref{eq:1.2}, we choose the orthonormal frame to be
\begin{equation}
\label{eq:4.4}
\begin{aligned}
&E^{\underline{u}}=\frac{1+f(r)}{2}du+dr,\quad E^{\underline{r}}=\frac{1-f(r)}{2}du-dr,\\
&E^{\underline{x}}=\sqrt{h(r)} dx.
\end{aligned}
\end{equation}
for which
\begin{equation}
\label{eq:4.5}
ds^2=\eta_{ab}E^aE^b,\quad \eta_{ab}={\rm diag}(-1,1,\dots,1).
\end{equation}
The spin connections for this frame are given by
\begin{equation}
\label{eq:4.6}
\begin{aligned}
&\omega_{\underline{ur}}=\frac{f'(r)}{2},\quad \omega_{\underline{ux}}=\frac{(-1+f(r))h'(r)}{4\sqrt{h(r)}},\\ &\omega_{\underline{rx}}=\frac{(1+f(r))h'(r)}{4\sqrt{h(r)}}.
\end{aligned}
\end{equation}
The form of gamma matrices in different dimensions is different~\cite{N1,Wilczek,Pethybridge}. We list the representations of gamma matrices for the 3-dimensional case as follows:
\begin{equation}
\label{eq:4.7}
\Gamma^{\underline{u}}= i \sigma^2,\quad \Gamma^{\underline{r}}=\sigma^3,\quad \Gamma^{\underline{x}}=\sigma^1.
\end{equation}
By inserting spinor $\psi_{\pm}(u,x,r)$ into~\eqref{eq:4.3} and using the gamma matrices defined in~\eqref{eq:4.7}, we will obtain two equations:
\begin{subequations}
	\begin{numcases}{}
\bigg(-2m-2i\omega-\frac{h'(r)}{4h(r)}-\frac{f'(r)}{2}-\frac{f(r)h'(r)}{4h(r)}\bigg)\notag\\
\times\psi_+(r)+\bigg(\frac{2ik}{\sqrt{h(r)}}-2i\omega+\frac{h'(r)}{4h(r)}-\frac{f'(r)}{2}\notag\\
-\frac{f(r)h'(r)}{4h(r)}\bigg)\psi_-(r)+\bigg(1-f(r)\bigg)\psi'_-(r)\notag\\
-\bigg(1+f(r)\bigg)\psi'_+(r)=0,\label{eq:4.8}\\
\bigg(-2m+2i\omega+\frac{h'(r)}{4h(r)}+\frac{f'(r)}{2}+\frac{f(r)h'(r)}{4h(r)}\bigg)\notag\\
\times\psi_-(r)+\bigg(\frac{2ik}{\sqrt{h(r)}}+2i\omega-\frac{h'(r)}{4h(r)}+\frac{f'(r)}{2}\notag\\
+\frac{f(r)h'(r)}{4h(r)}\bigg)\psi_+(r)+\bigg(1+f(r)\bigg)\psi'_-(r)\notag\\
-\bigg(1-f(r)\bigg)\psi'_+(r)=0.\label{eq:4.9}
\end{numcases}
\end{subequations}
We combine the two equations of~\eqref{eq:4.8} and~\eqref{eq:4.9}, then we expand them near the horizon $r=r_c$. The first-order equation near the horizon is
\begin{equation}
\begin{aligned}
\label{eq:4.10}
1st:\quad&\bigg[2i\omega+m-2\pi T+\frac{ik}{\sqrt{h(r_c)}}\bigg]\psi_+\\
&+\bigg[2i\omega-m-2\pi T-\frac{ik}{\sqrt{h(r_c)}}\bigg]\psi_-=0.
\end{aligned}
\end{equation}
We take the coefficients $\big(2i\omega+m-2\pi T+ik/\sqrt{h(r_c)}\big)$ and $\big(2i\omega-m-2\pi T-ik/\sqrt{h(r_c)}\big)$ to be 0 and thus there are two independent free parameters $\psi_+$ and $\psi_-$ to this equation. The first-order pole-skipping point is obtained as
\begin{equation}
\label{eq:4.11}
\omega_{\star}=-i\pi T, \quad k_{\star}=im\sqrt{h(r_c)}.
\end{equation}
We expand the Dirac equation to higher orders. For convenience, we use $h(r)=r^2$ when solving high-order equations:
\begin{subequations}
\begin{numcases}{}
2nd:\frac{1}{4}\bigg(\frac{4(m+\pi T-3ik\pi T)(m+2ik\pi T)}{m+2\pi T+2ik\pi T-2i\omega}+\frac{m^2}{\pi T}\notag\\
-2i\omega+6\pi T+4ik\pi T+4k^2\pi T\bigg)\psi^{0}_+\notag\\
+\frac{1}{2}\bigg(3-\frac{i\omega}{\pi T}\bigg)\psi^{1}_+=0,\label{eq:4.12}\\
3rd:\bigg(im^5+4m^3\pi T(10i\pi T+2ik^2\pi T+\omega)\notag\\
+2m^4(5i\pi T-k\pi T+\omega)+4m\pi^2T^2\notag\\
\times\big((81i+40k+64ik^2+4ik^4)\pi^2 T^2\notag\\
+4(7-2ik+3k^2)\pi T\omega-3i\omega^2\big)\notag\\
+8m^2\pi T\big((9+3ik+2k^2)\pi T\omega\notag\\
+2(10i-7k+5ik^2-k^3)\pi^2T^2-i\omega^2\big)\notag\\
-8\pi^2T^2\big(3i(9+5ik+4k^2)\pi T\omega^2\notag\\
-(69+100ik+92k^2+20ik^3+4k^4)\pi^2T^2\omega\notag\\
+(141k-45i-160ik^2+80k^3-20ik^4+4k^5)\pi^3T^3\notag\\
+2\omega^3\big)\bigg)\frac{\psi^{0}_+}{8\pi T(2k\pi T-2i\pi T-2\omega-im)(3\pi T-i\omega)}\notag\\
+\big(5-\frac{i\omega}{\pi T}\big)\psi^{2}_+=0,\label{eq:4.13}\\
\qquad \qquad \vdots\notag
\end{numcases}
\end{subequations}
The all-order pole-skipping points are
\begin{equation}
\label{eq:4.14}
\begin{aligned}
\omega_{\star}=-i\pi T, \quad k_{\star}=&\,\frac{im}{2\pi T}\,;\\	
\omega_{\star}=-3i\pi T, \quad k_{\star}=&\,-\frac{im}{2\pi T},\;\frac{im}{2\pi T}\pm 1\,;\\
\omega_{\star}=-5i\pi T, \quad k_{\star}=&\,\frac{im}{2\pi T},\;-\frac{im}{2\pi T}\pm 1,\;\frac{im}{2\pi T}\pm 2\,;\\
\omega_{\star}=-7i\pi T, \quad k_{\star}=&\,-\frac{im}{2\pi T},\;\frac{im}{2\pi T}\pm 1,\;-\frac{im}{2\pi T}\pm 2,\\
&\frac{im}{2\pi T}\pm 3\,;\\
\qquad \qquad \vdots
\end{aligned}
\end{equation}	
The operator dimension of the Dirac field in de Sitter spacetime is $\Delta=\frac{d-1}{2}+ imL$~\cite{Pethybridge}. The pole-skipping points~\eqref{eq:4.14} can be rewritten in terms of operator dimensions:
\begin{equation}
\label{eq:4.15}
\begin{aligned}
\omega_{\star} =-i\pi T, \quad k_{\star}=&\,\Delta-1\,;\\
\omega_{\star} =-3i\pi T, \quad k_{\star}=&\,\Delta-2,\;\Delta,\;-\Delta+1\,;\\
\omega_{\star} =-5i\pi T, \quad k_{\star}=&\,\Delta-3,\;\Delta-1,\;\Delta+1,\;-\Delta,\;-\Delta+2\,;\\
\omega_{\star} =-7i\pi T, \quad k_{\star}=&\,\Delta-4,\;\Delta-2,\;\Delta,\;\Delta+2,\;-\Delta-1,\\
&-\Delta+1,\;-\Delta+3\,;\\
&\qquad \qquad \vdots
\end{aligned}
\end{equation}
The pole-skipping of the Dirac field for BTZ black holes in AdS is given as~\cite{BlakeDavison}:
\begin{equation}
\label{eq:4.16}
\begin{aligned}
\omega_{\star} =-i\pi T, \quad k_{\star}=&\,i(\Delta-1)\,;\\
\omega_{\star} =-3i\pi T, \quad k_{\star}=&\,i(\Delta-2),\;i\Delta,\;-i(\Delta-1)\,;\\
\omega_{\star} =-5i\pi T, \quad k_{\star}=&\,i(\Delta-3),\;i(\Delta-1),\;i(\Delta+1),\;-i\Delta,\\
&-i(\Delta-2)\,;\\
\omega_{\star} =-7i\pi T, \quad k_{\star}=&\,i(\Delta-4),\;i(\Delta-2),\;i\Delta,\;i(\Delta+2),\\
&-i(\Delta+1),\;-i(\Delta-1),\;-i(\Delta-3)\,;\\
&\qquad \qquad \vdots
\end{aligned}
\end{equation}
From Eq.~\eqref{eq:4.15} and~\eqref{eq:4.16}, it can be observed that the frequencies of pole-skipping points in dS and AdS are also identical, and the momenta are also complex conjugates of each other in terms of the operator dimension. The Dirac field is a first-order differential equation. Unlike the second-order differential equation mentioned earlier, the mass in the exponential equation of the Dirac equation under the dS metric changes from $m$ to $im$; therefore, the operator dimension changes from $\Delta=\frac{d-1}{2}+mL$ in AdS to $\Delta=\frac{d-1}{2}+imL$ in dS~\cite{Pethybridge}. In the exponential equation, the momentum has also changed from $k$ to $ik$. This results in the momenta of the pole-skipping points in dS and AdS being complex conjugate to each other in terms of the operator dimension.

\subsection{Rarita-Schwinger field}
\label{sec:5}
We consider the Rarita-Schwinger field with spin-3/2 in dS$_4$ in this section. The metric we used here is also described in~\eqref{eq:1.2}. The action describing the massive Rarita-Schwinger field $\Psi_M$ is given by~\cite{N2,Volovich,Corley,Koshelev,Rashkov,Matlock}
\begin{equation}
\label{eq:5.1}
S_{RS}\propto\int d^{d}x\sqrt{-g}\bar{\Psi}_M(\Gamma^{MNP}\nabla_N-m\Gamma^{MP})\Psi_P.
\end{equation}
The covariant derivative acting on the spin-$\frac{3}{2}$ field is given by $\nabla_M\Psi_P=\partial_M\Psi_P-\tilde{\Gamma}^N_{MP}\Psi_N+\frac{1}{4}(\omega_{ab})_M\Gamma^{ab}\Psi_P$, where $\tilde{\Gamma}^N_{MP}$ is the Christoel symbol, and $\omega_M$ is the spin connection one form.  $\Gamma_a$ are Gamma matrices which satisfy Grassman algebra $\{\Gamma^a,\Gamma^b\}=2\eta^{ab}$, where $\eta^{ab}={\rm diag}(-1,+1,\dots,+1)$~\cite{Wilczek,Pethybridge}. The equation of motion derived from~\eqref{eq:5.1} is given as~\cite{N2,Volovich,Corley,Koshelev,Rashkov,Matlock}
\begin{equation}
\label{eq:5.2}
\Gamma^{MNP}\nabla_N \Psi_P-m\Gamma^{MN} \Psi_N=0.
\end{equation}
Since the background metric is vacuum, it can be proven from the Einstein equation that the above equation of motion is equivalent to the following formula~\cite{N2}
\begin{equation}
\label{eq:5.3}
(\Gamma^M\nabla_M+m)\Psi_N=0,
\end{equation}
with additional constraints
\begin{equation}
\label{eq:5.4}
\Gamma^M\Psi_M=0,\quad \nabla^M\Psi_M=0.
\end{equation}
We choose the orthonormal frame to be
\begin{equation}
\label{eq:5.5}
\begin{aligned}
&E^{\underline{u}}=\frac{1+f(r)}{2}du+dr,\quad E^{\underline{r}}=\frac{1-f(r)}{2}du-dr,\\
&E^{\underline{x_\alpha}}=r dx_\alpha,
\end{aligned}
\end{equation}
for which
\begin{equation}
\label{eq:5.6}
ds^2=\eta_{ab}E^aE^b,\quad \eta_{ab}={\rm diag}(-1,1,\dots,1).
\end{equation}
The spin connections for this frame are given by
\begin{equation}
\label{eq:5.7}
\begin{aligned}
	&\omega_{\underline{ur}}=\frac{f'(r)}{2},\quad \omega_{\underline{ux_\alpha}}=\frac{(-1+f(r))h'(r)}{4\sqrt{h(r)}},\\ &\omega_{\underline{rx_\alpha}}=\frac{(1+f(r))h'(r)}{4\sqrt{h(r)}}.
\end{aligned}
\end{equation}
The form of gamma matrices in different dimensions is different~\cite{N1,Wilczek,Pethybridge}. We list the representations of gamma matrices for the 4-dimensional case as follows:
\begin{equation}
\label{eq:5.8}
\Gamma^{\underline{u}}=\sigma^1\otimes i \sigma^2,\; \Gamma^{\underline{r}}=\sigma^3\otimes \mathbbm{1},\; \Gamma^{\underline{x_1}}=\sigma^1\otimes \sigma^1,\; \Gamma^{\underline{x_2}}=\sigma^1\otimes \sigma^3.
\end{equation}
the metric components depend only on the $r$ coordinate and the plane wave is $\Psi_M(u,r,x^\alpha)=\Psi_M(r)e^{-i\omega u+ik_\alpha x^\alpha}$. The Rarita-Schwinger field $\Psi_M(r)$ have four components $(M=u,r,x_1,x_2)$, and all vector components can be decomposed as~\cite{N2}
\begin{equation}
\label{eq:5.9}
\Psi_M(r)=\sum_{x_1=\pm}\sum_{x_2=\pm}\Psi_M^{(x_1,x_2)}(r)
\end{equation}
with $x_{1,2}=\pm$. Each of the components in the decomposition contains a quarter of the total degrees of freedom of the spinor. By considering the $uu-$component of the dual bulk excitation~\cite{Grozdanov1,Blake1,Blake2}, the leading order pole-skipping point of the energy density correlation function can be identified. Therefore, we will also proceed by obtaining the leading pole-skipping point by just considering the $u-$component of the Rarita-Schwinger field:
\begin{equation}
\label{eq:5.10}
\begin{aligned}
(\Gamma^M D_M+m)\Psi_u=\frac{\partial_rf(r)}{2}\bigg\{-\big[\Gamma^{\underline{u}}+\Gamma^{\underline{r}}\big]\Psi_u\\
+\frac{1}{2}\big[\big(1+f(r)\big)\Gamma^{\underline{u}}-\big(1-f(r)\big)\Gamma^{\underline{r}}\big]\Psi_r\bigg\}.
\end{aligned}
\end{equation}
By using the Gamma matrices~\eqref{eq:5.8}, Eq.~\eqref{eq:5.10} becomes
\begin{subequations}
	\begin{numcases}{}
		\big((1-f(r))h'(r)-(4i\omega+f'(r))h(r)\notag\\
		+4ik\sqrt{h(r)}\big)\Psi^{(-,-)}_u+\big((4m-4i\omega-f'(r))h(r)\notag\\
		-(1+f(r))h'(r)\big)\Psi^{(+,+)}_u+2h(r)\big(1-f(r)\big)\notag\\
		\times\partial_r\Psi^{(-,-)}_u-2h(r)\big(1+f(r)\big)\partial_r\Psi^{(+,+)}_u+2h(r)\notag\\
		\times f'(r)\Gamma^{\underline{u}}\big(\Psi_u-\frac{1}{2}\big(1+f(r)\big)\Psi_r\big)+2h(r)f'(r)\notag\\
		\times \Gamma^{\underline{r}}\big(\Psi_u+\frac{1}{2}\big(1-f(r)\big)\Psi_r\big)=0,\label{eq:5.11}\\
		\big((4i\omega+f'(r))h(r)-(1-f(r))h'(r)\notag\\
		+4ik\sqrt{h(r)}\big)\Psi^{(+,+)}_u+\big((4m+4i\omega+f'(r))h(r)\notag\\
		+(1+f(r))h'(r)\big)\Psi^{(-,-)}_u-2h(r)\big(1-f(r)\big)\notag\\
		\times\partial_r\Psi^{(+,+)}_u+2h(r)\big(1+f(r)\big)\partial_r\Psi^{(-,-)}_u+2h(r)\notag\\
		\times f'(r)\Gamma^{\underline{u}}\big(\Psi_u-\frac{1}{2}\big(1+f(r)\big)\Psi_r\big)+2h(r)f'(r)\notag\\
		\times \Gamma^{\underline{r}}\big(\Psi_u+\frac{1}{2}\big(1-f(r)\big)\Psi_r\big)=0.\label{eq:5.12}
	\end{numcases}
\end{subequations}
Expanding equations Eq.~\eqref{eq:5.11} and Eq.~\eqref{eq:5.12} near the cosmic horizon $r_c=L$, we could obtain the first-order equation near the horizon:
\begin{equation}
	\begin{aligned}
		\label{eq:5.13}
	1st:\quad&\bigg[2\pi T+m+2i\omega-\frac{ik}{\sqrt{h(r_c)}}\bigg]\Psi^{(-,-)}_{u,0}\\
		&+\bigg[2\pi T-m+2i\omega+\frac{ik}{\sqrt{h(r_c)}}\bigg]\Psi^{(+,+)}_{u,0}=0.
	\end{aligned}
\end{equation}
We take the coefficients $\big(2\pi T+m+2i\omega-ik/\sqrt{h(r_c)}\big)$ and $\big(2\pi T-m+2i\omega+ik/\sqrt{h(r_c)}\big)$ to be 0 and thus there are two independent free parameters $\Psi^{(-,-)}_{u,0}$ and $\Psi^{(+,+)}_{u,0}$ to this equation. The first-order pole-skipping point is obtained as
\begin{equation}
	\label{eq:5.14}
	\omega_{\star}=i\pi T, \quad k_{\star}=-im\sqrt{h(r_c)}.
\end{equation}
We could also consider $r$ component of the Rarita-Schwinger field
\begin{equation}
	\label{eq:5.15}
	\begin{aligned}
(\Gamma^M D_M+m)\Psi_r=\frac{\partial_rf(r)}{2}\big[\Gamma^{\underline{u}}+\Gamma^{\underline{r}}\big]\Psi_r.
\end{aligned}
\end{equation}
By using the Gamma matrices~\eqref{eq:5.8}, Eq.~\eqref{eq:5.15} becomes
\begin{subequations}
	\begin{numcases}{}
		\big((1-f(r))h'(r)-(4i\omega+f'(r))h(r)\notag\\
		+4ik\sqrt{h(r)}\big)\Psi^{(-,-)}_r+\big((4m-4i\omega-f'(r))h(r)\notag\\
		-(1+f(r))h'(r)\big)\Psi^{(+,+)}_r+2h(r)\big(1-f(r)\big)\notag\\
		\times\partial_r\Psi^{(-,-)}_r-2h(r)\big(1+f(r)\big)\partial_r\Psi^{(+,+)}_r\notag\\
		-2h(r)f'(r)(\Gamma^{\underline{v}}+\Gamma^{\underline{r}})\Psi_r=0,\label{eq:5.16}\\
		\big((4i\omega+f'(r))h(r)-(1-f(r))h'(r)\notag\\
		+4ik\sqrt{h(r)}\big)\Psi^{(+,+)}_r+\big((4m+4i\omega+f'(r))h(r)\notag\\
		+(1+f(r))h'(r)\big)\Psi^{(-,-)}_r-2h(r)\big(1-f(r)\big)\notag\\
		\times\partial_r\Psi^{(+,+)}_r+2h(r)\big(1+f(r)\big)\partial_r\Psi^{(-,-)}_r\notag\\
			-2h(r)f'(r)(\Gamma^{\underline{v}}+\Gamma^{\underline{r}})\Psi_r=0.\label{eq:5.17}
	\end{numcases}
\end{subequations}
Expanding equations Eq.~\eqref{eq:5.16} and Eq.~\eqref{eq:5.17} near the cosmic horizon $r_c=L$, we could obtain the first-order equation near the horizon:
\begin{equation}
	\begin{aligned}
		\label{eq:5.18}
	1st:\quad&\bigg[2i\omega-6\pi T+m-\frac{ik}{\sqrt{h(r_c)}}\bigg]\Psi^{(-,-)}_{r,0}\\
		&+\bigg[2i\omega-6\pi T-m+\frac{ik}{\sqrt{h(r_c)}}\bigg]\Psi^{(+,+)}_{r,0}=0.
	\end{aligned}
\end{equation}
We take the coefficients $\big(2i\omega-6\pi T+m-ik/\sqrt{h(r_c)}\big)$ and $\big(2i\omega-6\pi T-m+ik/\sqrt{h(r_c)}\big)$ to be 0 and thus there are two independent free parameters $\Psi^{(-,-)}_{r,0}$ and $\Psi^{(+,+)}_{r,0}$ to this equation. The first-order pole-skipping point obtained from $r$ component is
\begin{equation}
	\label{eq:5.19}
	\omega_{\star}=-3i\pi T, \quad k_{\star}=-im\sqrt{h(r_c)}.
\end{equation}
We express results~\eqref{eq:5.14} and~\eqref{eq:5.19} in the form of the operator dimension. Analogy to the situation in the operator dimension of the Dirac field transitioning from AdS spacetime to dS spacetime, we adopt $\Delta=\frac{d-1}{2}+imL$ to represent the operator dimension of the Rarita-Schwinger field in dS spacetime. For convenience, we choose $h(r)=r^2$. The Eq.~\eqref{eq:5.14} and Eq.~\eqref{eq:5.19} become 
\begin{equation}
	\label{eq:5.20}
	\omega_{\star}=i\pi T, \quad k_{\star}=1-\Delta,
\end{equation}
and
\begin{equation}
	\label{eq:5.21}
	\omega_{\star}=-3i\pi T, \quad k_{\star}=1-\Delta,
\end{equation}
Compare these two results,~\eqref{eq:5.20} and~\eqref{eq:5.21}, with the leading-order pole-skipping of Rarita-Schwinger field in AdS spacetime~\cite{N2}:
\begin{equation}
	\label{eq:5.22}
	\omega_{\star}=i\pi T, \quad k_{\star}=i(1-\Delta),
\end{equation}
and
\begin{equation}
	\label{eq:5.23}
	\omega_{\star}=-3i\pi T, \quad k_{\star}=i(1-\Delta).
\end{equation}
This situation is analogous to that of the Dirac field. Compared to the mass and momentum in the exponential equation in AdS, the mass and momentum in the exponential equation in dS will change from $m$ to $im$ and from $k$ to $ik$, respectively. These changes will lead to a transformation in the mass component of the operator dimensions: $\Delta=\frac{d-1}{2}+mL\rightarrow\Delta=\frac{d-1}{2}+imL$, as well as a shift in the momenta at the pole-skipping points, $k_\star\rightarrow ik_\star$, in terms of the operator dimension.

\section{Conclusion}
\label{sec:Conclusion}
In this paper, we consider the higher-order pole-skipping points of the scalar and the vector field, as well as the pole-skipping points of the fermionic field, within the dS horizon structure. Our result indicates that when selecting the incoming wave condition at the cosmic horizon $r=r_c=L$ in dS spacetime, the pole-skipping frequencies of the bosonic and fermionic fields are the same as those when selecting the incoming wave condition at the AdS horizon, that is, for different spin $s$ fields, $\omega=i2\pi T(s-1)$ is satisfied.

If we perform a coordinate transformation on the dS metric~\eqref{eq:1.1} as $z=L^2/r$, and compare the resulting exponential equation as $z$ approaches $0$ with the corresponding exponential equation of the AdS metric in the coordinate $z$, we find that for both bosonic and fermionic fields, the mass $m$ and momentum $k$ are complex conjugate to each other. We list the changes in the mass dimension relations from the AdS/CFT correspondence to the dS/CFT correspondence in TABLE~\ref{6.1}.
\begin{table}[h!]
\begin{center}
\caption{$d-$dimensional (A)dS/CFT Mass Dimension Relations}
\label{6.1}
\begin{tabular}{c|c|c}
\textbf{spin fields}&\textbf{AdS/CFT}&\textbf{dS/CFT}\\
\hline
Scalar & $\Delta(\Delta+1-d)=m^2L^2$ & $\Delta(\Delta+1-d)=-m^2L^2$\\
Maxwell & $(\Delta-1)(\Delta+2-d)=0$ & $(\Delta-1)(\Delta+2-d)=0$\\
Gravity & $\Delta(\Delta+1-d)=0$ & $\Delta(\Delta+1-d)=0$\\
Dirac & $\Delta-\frac{d-1}{2}=mL$ & $\Delta-\frac{d-1}{2}=imL$\\
R-S & $\Delta-\frac{d-1}{2}=mL$ & $\Delta-\frac{d-1}{2}=imL$\\
\end{tabular}
\end{center}
\end{table}
In previous studies, only the mass transition $m\rightarrow im$ from AdS/CFT to dS/CFT, causing the transition of operator dimension $\Delta$ in the mass component, was considered. In this paper, we show that opposite changes in spacetime curvature can lead to opposite values of momenta on the imaginary axis $k_\star\rightarrow ik_\star$ at the pole-skipping points in measure of the operator dimension. We hope that those variations in mass and momentum exhibited by the pole-skipping phenomenon could provide an exploration for the holographic dual theory of dS spacetime.

\section*{Acknowledgments}
This work is supported by the National Natural Science Foundation of China (No.12275166), Natural Science Foundation of Henan Province of China (No.252300420892), and the Launching Funding of Henan University of Technology (No.31401598). This work was also supported by the Basic Science Research Program through the National Research Foundation of Korea (NRF) funded by the Ministry of Science, ICT and Future Planning (NRF-2021R1A2C1006791), the GIST Research Institute (GRI) and the AI-based GIST Research Scientist Project grant funded by the GIST in 2023.

\appendix
\section{Details of near-horizon expansions of the scalar field} 
\label{sec:Details1}
In this appendix, we show the details of the near-horizon expansions of the scalar field. We can calculate a Taylor series solution to the scalar field $\Phi(r)$ equation of motion when the matrix equation~\eqref{eq:1.15} is satisfied. The first few elements of this matrix are shown below
\begin{equation}
		\begin{aligned}
M_{11}&=\frac{k^2+m^2r^2_c-i\omega r_c}{2r_c^2};\\
M_{21}&=\frac{2m^2r_c-i\omega}{4r_c^2},\\
M_{22}&=\frac{k^2+m^2r^2_c+12\pi Tr_c-5i\omega r_c-r_c^2f''(r_c)}{4r_c^2};\\
M_{31}&=\frac{m^2}{6r^2_c},\\
M_{32}&=\frac{k^2+m^2r^2_c+40\pi Tr_c-9i\omega r_c-3r^2_cf''(r_c)}{12r^2_c},\\
M_{33}&=\frac{4m^2r_c+16\pi T-6i\omega-5r_cf''(r_c)-r^2_cf^{(3)}(r_c)}{6r^2_c};\notag
	\end{aligned}
\end{equation}
\begin{equation}
		\begin{aligned}
M_{41}&=0,\\
M_{42}&=\frac{6m^2-9f''(r_c)-7r_cf^{(3)}(r_c)-r^2_cf^{(4)}(r_c)}{48r^2_c},\\
M_{43}&=\frac{6m^2r_c+72\pi T-15i\omega-21r_cf''(r_c)-4r^2_cf^{(3)}(r_c)}{24r^2_c},\\
M_{44}&=\frac{k^2+m^2r^2_c+84\pi Tr_c-13i\omega r_c-6r^2_cf''(r_c)}{8r^2_c};\\
M_{51}&=0,\\
M_{52}&=\frac{-16f^{(3)}(r_c)-9r_cf^{(4)}(r_c)-r^2_cf^{(5)}(r_c)}{240r^2_c},\\
M_{53}&=\frac{12m^2-48f''(r_c)-36r_cf^{(3)}(r_c)-5r^2_cf^{(4)}(r_c)}{120r^2_c},\\
M_{54}&=\frac{8m^2r_c+192\pi T-28i\omega-54r_cf''(r_c)-10r^2_cf^{(3)}(r_c)}{40r^2_c},\\
M_{55}&=\frac{k^2+m^2r^2_c+144\pi Tr_c-17i\omega r_c-10r^2_cf''(r_c)}{10r^2_c};\\
\qquad \vdots
	\end{aligned}
\end{equation}

\section{Details of near-horizon expansions of the Maxwell field} 
\label{sec:Details2}
In this appendix, we show the details of the near-horizon expansions of the Maxwell field. When we use a method similar to that for calculating higher-order pole-skipping points in the scalar field, we can derive a Taylor series solution to the vector part of the Maxwell field $A_y(r)$ equation of motion~\eqref{eq:2.5}. The first few elements of the matrix are shown below
\begin{equation}
		\begin{aligned}
M_{11}&=\frac{k^2}{2r_c^2};\\
M_{21}&=0,\;M_{22}=\frac{k^2+8\pi Tr_c-4i\omega r_c-r_c^2f''(r_c)}{4r_c^2};\\
M_{31}&=0,\;M_{32}=\frac{8\pi T-4i\omega-4r_cf''(r_c)-r^2_cf^{(3)}(r_c)}{12r^2_c},\\
M_{33}&=\frac{k^2+32\pi T r_c-8i\omega r_c-3r^2_cf''(r_c)}{6r^2_c};\\
M_{41}&=0,\;M_{42}=-\frac{6f''(r_c)+6r_cf^{(3)}(r_c)+r^2_cf^{(4)}(r_c)}{48r^2_c},\\
M_{43}&=\frac{24\pi T-6i\omega-9r_cf''(r_c)-2r^2_cf^{(3)}(r_c)}{12r^2_c},\\
M_{44}&=\frac{k^2+72\pi Tr_c-12i\omega r_c-6r^2_cf''(r_c)}{8r^2_c};\\
M_{51}&=0,\;M_{52}=-\frac{12f^{(3)}(r_c)+8r_cf^{(4)}(r_c)+r^2_cf^{(5)}(r_c)}{240r^2_c},\\
M_{53}&=-\frac{32r_cf^{(3)}(r_c)+5r^2_cf^{(4)}(r_c)+36f''(r_c)}{120r^2_c},\\
M_{54}&=\frac{144\pi T-24i\omega-48r_cf''(r_c)-10r^2_cf^{(3)}(r_c)}{40r^2_c},\\
M_{55}&=\frac{k^2+128\pi Tr_c-16i\omega r_c-10r^2_cf''(r_c)}{10r^2_c};\\
\qquad \vdots
	\end{aligned}
\end{equation}

\end{document}